\documentclass[twocolumn,showpacs,preprintnumbers,amsmath,amssymb,floatfix]{revtex4}
\usepackage{graphicx}
\usepackage{epstopdf}

\usepackage[center]{subfigure}
\usepackage{bigints}
\usepackage{cancel}
\usepackage{amsmath}
\usepackage{mathbbol}
\usepackage[center]{subfigure}
\usepackage{tikz}
\usepackage{standalone}
\usepackage{mathtools}

\begin{document}
\author{Sara Najem $^{1}$ }
\affiliation{$^1$ National Center for Remote Sensing, National Council for Scientific Research (CNRS), Riad al Soloh, 1107 2260, Beirut, Lebanon}

%\affiliation{$^2$ Physics Department, Rutherford Building, 3600 rue
%University, McGill University, Montr\'eal, Qu\'ebec, Canada H3A 2T8}
%\affiliation{$^3$ Department of Ecology and Evolutionary Biology, Princeton University, Princeton, NJ, USA}

\date{\today}

\begin{abstract}
In this Letter we follow the short-ranged Syrian refugees' migration to Lebanon as documented by the UNHCR. We propose a model inspired by the Debye-H\"{u}ckel theory and show that it properly predicts the refugees' mobility while the gravity model fails. We claim that the interaction between origin cities attenuates and/or extenuates the flux to destinations, and consequently, in analogy with the effective charges of interacting particles in a plasma, these source cities are characterized by effective populations determined by their pairwise remoteness/closeness and defined by areas of control between the fighting parties.  
 \end{abstract}
%\pacs{89.75.-k, 89.65.Lm, 89.75.Kd, 88.40.fc}
\title{Debye-H\"{u}ckel Theory for Refugees' Migration}

\maketitle
The ubiquity of mobile phone usage data and credit card transactions made human mobility amenable to mathematical analysis, and therefore, lead to the discovery of underlying patterns of motion described as random walks and particularly L\'{e}vy flights \cite{Colak:2015es,brockmann:2006du,Schneider:2013jc}. It also revealed universal behavior, which was explained by the gravity and the radiation models  \cite{Masucci:2013fx,Yan:2014cb,Wang:2014gu,Song:2010kv,Gonzalez:2008hy,Simini:2012fe,2013JSP...151..304H}, which also describes trade flow \cite{Silva:2006th,Anderson:2003cn} among other types of traffic. The first links the flux between origin and destination cities to powers of their initial populations with an inverse dependence on their pairwise distance given by:

  \begin{equation}\label{firstscaling}
T_{ij} \propto \frac{m_i^{\alpha} n_j^{\beta}}{d_{ij}^{\gamma}},
\end{equation}
where $T_{ij}$ is the influx between the source city $i$ and the destination city $j$, $m_i$ and $n_j$ are their respective initial populations, $d_{ij}$ is their pairwise distance, $\alpha$, $\beta$, and $\gamma$ are the model parameters. On the other hand the radiation model is a parameter-free model given by:

  \begin{equation}\label{radiation }
T_{ij} = T_i \frac{m_i n_j}{(m_i + s_{ij}) (m_i+n_j + s_{ij})},
\end{equation}
where $T_i = \sum_{i \neq j} T_{ij}$ the total number of commuters at location $i$, and $s_{ij}$ is the population density in a circle centered at $i$. 
These two competing models have been shown to properly predict human mobility and their relative likelihood has been explored \cite{Simini:2012fe,barbosa2017human}. However, the question of their validity to model mobility under stressful conditions beyond the daily commute or travel has not been examined. 
Of particular interest is the refugee migration, which is a form of human mobility under life threatening circumstances.  Unlike the long timescale decision underlying the job selection process according to which the origin and destination are determined, refugees' migration is a pressing decision made on a very short timescale to flee a endangering environment. In this paper we are specifically concerned with the displaced Syrian population to neighboring countries and particularly to Lebanon, the most vulnerable country to the crisis spillover because of the large refugee  influx straining the infrastructure and further destabilizing Lebanon's socioeconomic conditions 
\cite{Anonymous:EX4Lti7C,Cherri:2016kd,Salem:2012vp}. The unavailability of census data and population densities in Lebanon restricted our data exploration and model fitting to a modification of the gravity model and made it impossible to test the explanatory potential of the parameter-free radiation model.
We thus present an adjusted gravity model incorporating the peculiarities of refugees' mobility using the United Nations Human Rights Council (UNHCR) data made available to our lab through an official data sharing request documenting their influx up till 2014 when the  Lebanese-Syrian borders were closed \cite{Cherri:2016kd}. The data chronicles the cumulative refugees' count until May 2014 detailing of the origin and destination cities. 

In econometrics, correction had been introduced to the gravity model as it failed to take into consideration the multilateral resistance terms in trade. The equation was either augmented with an exponential term accounting for the importer exporter difference, known as the Anderson–van Wincoop Gravity equation or with a stochastic effect \cite{Anderson:2003cn}. Physically this is equivalent to shifting from an isolated cities paradigm to a description that takes into account the cities interactions.
Analogously, Coulomb's law, a special case of  Eq. \ref{firstscaling} for $\alpha, \beta = 1$ and $\gamma = 2$, which describes the interaction between two isolated particles with charges $m_i$ and $n_j$ is invalidated when the particles are in an electrolyte or a plasma. More precisely, the surrounding particles screen the potential and subsequently their effect is equivalent  to the renormalization of the bare charge from $m_i \rightarrow m_i e^{-r/l}$, where $l$ is the Debye H\"{u}ckel length.

\begin{figure}[!htp] %
 % \includestandalone[scale=0.5]{scaling0}   
   \includegraphics[scale=0.3]{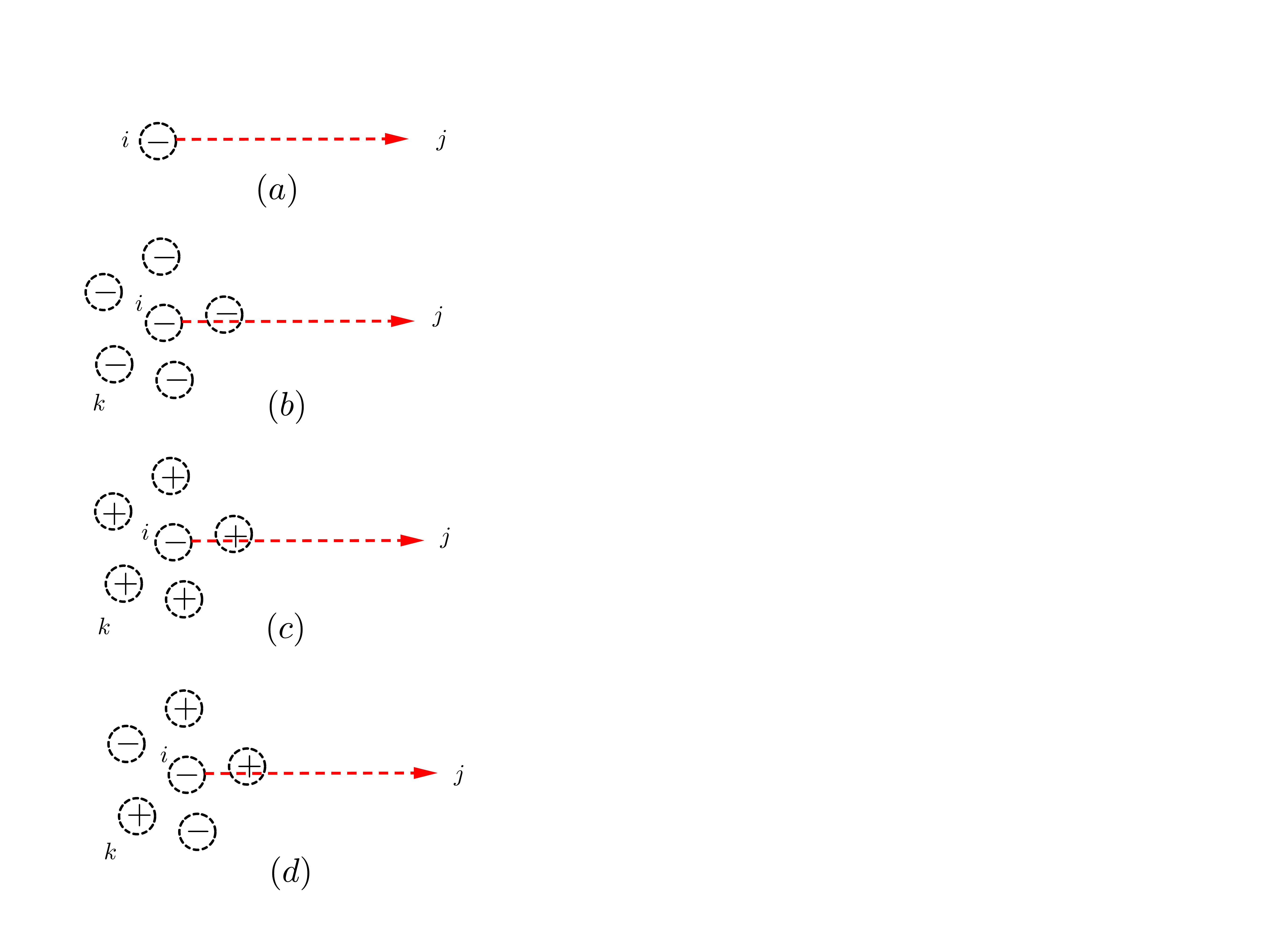}   
  %\vspace{2cm}
%  \includegraphics{Fig3.eps}
\caption{A scheme illustrating the force at $j$ exerted by an isolated particle shown in (a),  compared to the one in an electrolyte. Different configurations have different effects on the force at $j$.    }
\label{fig:renormforce} 
\end{figure}

Based on the above, two basic contentions underly our paper: when short range mobility between non isolated cities is studied, the interaction between the source cities cannot be neglected and this is the reason why the gravity model fails to explain the migration data. Thus, we argue that the latter is the ideal limit behavior and, therefore, accounting for the presence of multiple nearby source cities and their effect in shielding or accentuating the flux, allows to properly predict the migration as in Fig. \ref{fig:renormforce}, where a different shielding is generated by each configuration. 

Our second argument is based on the symmetry breaking nature of the refugees' migration. The flux becomes almost unidirectional as people from the host cities are unlikely to leave to unsafe destinations, that is: $T_{ij} >> T_{ji}$ even when their populations $m_i$ and $n_j$ are comparable, where $i$ and $j$ are  Syrian and  Lebanese cities respectively.    
 For this we rewrite the gravity model with the renormalized effect:%\nocite{oreg,schn,pond,smith,marg,hunn,advi,koha,mouse}

%%%%%%%%%%%%%%%%%%%%%%%%%%%%%%%%%%%%%%%%%%%%%%
%%                                          %%
%% Backmatter begins here                   %%
%%                                          %%
%%%%%%%%%%%%%%%%%%%%%%%%%%%%%%%%%%%%%%%%%%%%%%

  \begin{equation}\label{secondscaling}
T_{ij} = A \frac{m_i^{\alpha} n_j^{\beta}}{d_{ij}^{\gamma}}e^{\sum \limits_{k \in \mathcal{O}}  R_{ik}/\delta_k},
\end{equation}
where $R_{ik}$ is the distance between the origin city $i$ and the rest of the origins $k$ and $\delta_k$ is a model parameter. Each origin city is tagged with a corresponding $\delta_k$, which shields the flux when it is negative and accentuates it for positive values of the parameter as illustrated in Fig .\ref{fig:renormforce}. Therefore, $|\delta_k|$ can be thought of as a characteristic length marking the influence of the origin city $k$ on the migration from $i$, while the sign of $\delta_k$ determines the effective population of $k$.  

Moreover, Eq. \ref{secondscaling} incorporates the highly non-symmetrical nature of the flux. Explicitly, when $i$ denotes a Lebanese city and $j$ is a Syrian destination the exponential term introduces the possibility of obstructing  the flow in this direction for negative values of  $\delta_k$ .

We used Google API to calculate the cities pairwise distance, which is the shortest path over the road network. Additionally, in the face of Lebanon's lack of census data and the unavailability of official Syrian population data we retrieved their estimates from the City population website \cite{pop}. 

\begin{figure}[!htp]  
  \includegraphics[scale=0.4]{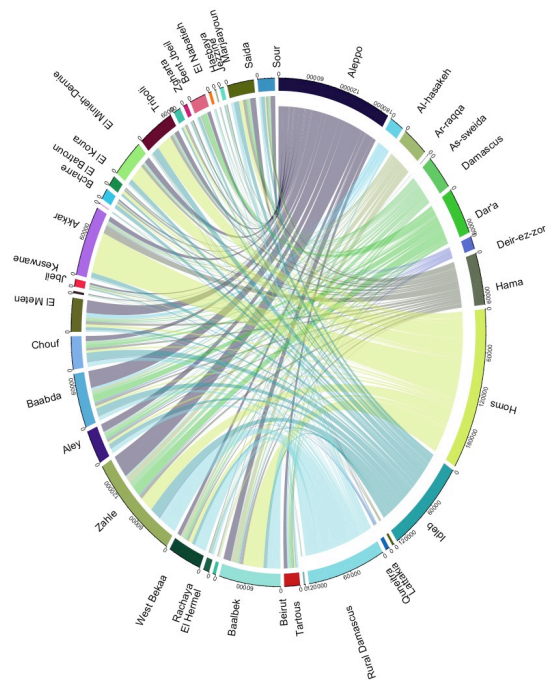} 
    \caption{The chord diagram depicts the flow between Syrian and Lebanese cities}
  \label{fig:chord}
 \end{figure}

\begin{figure}[!htp] %
 % \includestandalone[scale=0.5]{scaling0}   
   \includegraphics{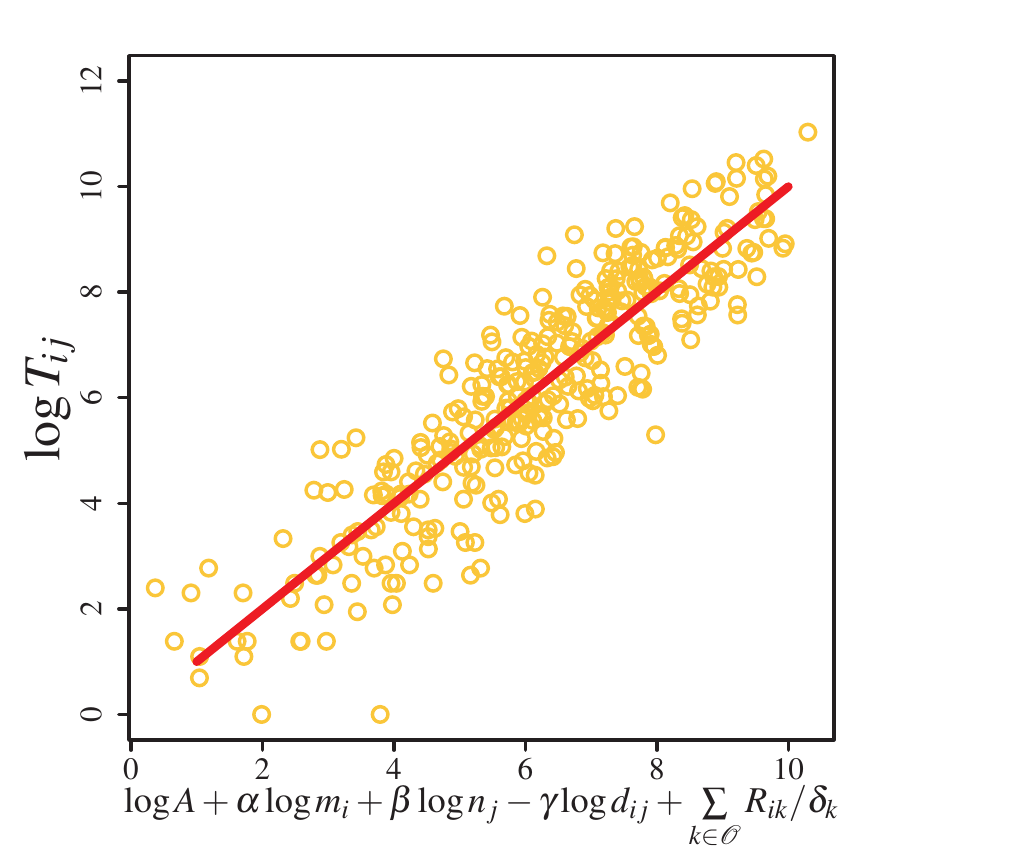}   
  %\vspace{2cm}
%  \includegraphics{Fig3.eps}
\caption{$\log{A} = 61 \pm 11$, $\alpha =  1.76 \pm 0.25$, $\beta = 1.13 \pm 0.06$, $\gamma = 2.21 \pm 0.20$ and $\delta_k$ are given in Table \ref{deltatable} and the Multiple R-squared:  0.8082.}
\label{fig:renormpotential} 
\end{figure}

\begin{figure}[!htp]  
  \includegraphics[scale=0.32]{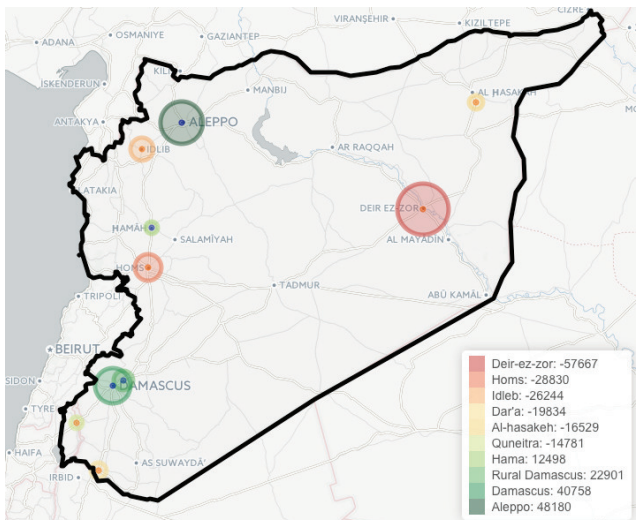} 
    \caption{The map shows the cities' centers with circumscribing circles whose radii are proportional to $1/\delta_k$ and whose colors are defined by the sign of $\delta_k$. Red and green represent negative and positive values of $\delta_k$ respectively. }
  \label{fig:map}
 \end{figure}

Figure \ref{fig:chord} shows the chord diagram depicting the flow from Syria to Lebanon with the width of the chords characterizing the flux intensity. Figure \ref{fig:renormpotential} shows the UNHCR data together with the red curve fitting the logarithm of Eq. \ref{secondscaling} given by $\log{T_{ij}}$ as a function of the $\log{A} + \alpha \log{m_i} + \beta \log{n_j} - \gamma \log{d_{ij}} + \sum_{k \in \mathcal{O}} R_{ik}/ \delta _k$ with a multiple R-squared of 0.8082. Table \ref{deltatable}  summarizes the values of $\delta_k$ while Fig. \ref{fig:map} shows the map of Syria with circles centered at the source cities' with radii proportional to $1/\delta_k$ and color coded by the sign of their corresponding $\delta_k$. 

 %\newline
 
 \begin{table}[h!]
  \centering
%\caption{My caption}
\label{deltatable}
\begin{tabular}{@{}ll@{}}
%\toprule
\multicolumn{1}{c}{$1/\l_k$ [\textit{$m^{-1}$}]}  & City        \\     %\\ \midrule
-57667   $\pm 7388$    & Deir-ez-zor    \\
-28830 $\pm 5741$      & Homs           \\
-26243 $\pm 2117$        & Idleb          \\
-19834 $\pm 4018$      & Dar'a          \\
-16529 $\pm 2271$       & Al-hasakeh     \\
-14781 $ \pm 1354$       & Quneitra       \\
12497 $\pm 2121$       & Hama           \\
22901 $ \pm 4809$        & Rural Damascus \\
40758 $\pm 14272$         & Damascus       \\
48180 $ \pm 7530$       & Aleppo       %  \\ \bottomrule
\end{tabular}
\caption{The table shows the values of $\delta_k$ corresponding to each Syrian city.}

\end{table}
 \newpage
The results of our model revealed the existence of two types of origin cities depending on the sign of their corresponding $\delta_k$. 
We suspect that this is the result of the interplay between areas controlled by the central government and areas controlled by the rebels, and is thus equivalent to an internally drawn border constraining migration and defining its routes.  
To check this hypothesis we compared our findings with reports delineating areas of control and show that our findings are strikingly in agreement with these maps \cite{fabrice}.  In the latter, cities under central government control are shown in light purple and these correspond to the cities with positive $\delta_k$ in our model, which we denote by $\mathcal{R}$, while the others held by all the regime's opponents correspond to cities with negative $\delta_k$ denoted by $\mathcal{RO}$. 
Particularly, when people are far from a region $\mathcal{R}$ under government control they are more likely to migrate while the opposite effect occurs when they are far from $\mathcal{RO}$. More precisely, when $R_{ik}/{\delta_k} >>0$, that is when the origin city $i$ is very far from $k$, where the latter is under government control the migration out of $i$ intensifies. Conversely, if a city $i$ is distant from $k$ where the latter is under $\mathcal{RO}$ the likelihood of migration decreases. Thus, the cities with positive $\delta_k$ play the role of attractors for the migration out of the source cities, while those with negative $\delta_k$ have the contrary effect. Therefore, our parameter $\delta_k$ is a proxy to understanding how the local population perceives the safety and security of their origin cities in relation to their remoteness from $\mathcal{R}$ and $\mathcal{RO}$.

In this paper, we have presented a model for refugees' migration, which is based on the idea of interaction between source cities. Effectively, this resulted in a renormalization in the source cities' population as our model suggested.  Consequently, cities were  classified according to the sign of their corresponding $\delta_k$. This sign difference was linked to the interplay between different areas of control leading to a space dependent migration undergoing varying degrees of friction.  Our model thus represents an attempt to predicting human mobility in relation to space and its fragmentation between the fighting parties. The analysis of these patterns should also complemented with migration data to other neighboring countries, which we did not have access to.

\section*{Acknowledgements}
We acknowledge the Director of the National Center for Remote Sensing Dr. Ghaleb Faour's role in officially acquiring the dataset from UNHCR. 
\section*{References}
%\newpage
\bibliography{bmc_article}
\end{document}